\batchmode
\makeatletter
\def\input@path{{\string"E:/Trabajo Angel/Mis articulos/Finished/Fusion 2019/Spooky effect OSPA/\string"}}
\makeatother
\documentclass[british,english]{IEEEtran}
\usepackage[T1]{fontenc}
\usepackage[latin9]{inputenc}
\usepackage{amsmath}
\usepackage{amsthm}
\usepackage{amssymb}
\usepackage{stmaryrd}
\usepackage{graphicx}

\makeatletter
\theoremstyle{plain}
\newtheorem{thm}{\protect\theoremname}
\theoremstyle{definition}
\newtheorem{defn}[thm]{\protect\definitionname}
\theoremstyle{plain}
\newtheorem{prop}[thm]{\protect\propositionname}
\theoremstyle{plain}
\newtheorem{lem}[thm]{\protect\lemmaname}
\theoremstyle{definition}
\newtheorem{example}[thm]{\protect\examplename}

\usepackage{cite} 
\usepackage[margin=8pt,font=footnotesize]{caption}
\usepackage{algorithm}
\usepackage{algpseudocode}

\usepackage{amsmath}  
\allowdisplaybreaks

\@ifundefined{showcaptionsetup}{}{%
 \PassOptionsToPackage{caption=false}{subfig}}
\usepackage{subfig}
\makeatother

\usepackage{babel}
\addto\captionsbritish{\renewcommand{\definitionname}{Definition}}
\addto\captionsbritish{\renewcommand{\examplename}{Example}}
\addto\captionsbritish{\renewcommand{\lemmaname}{Lemma}}
\addto\captionsbritish{\renewcommand{\propositionname}{Proposition}}
\addto\captionsbritish{\renewcommand{\theoremname}{Theorem}}
\addto\captionsenglish{\renewcommand{\definitionname}{Definition}}
\addto\captionsenglish{\renewcommand{\examplename}{Example}}
\addto\captionsenglish{\renewcommand{\lemmaname}{Lemma}}
\addto\captionsenglish{\renewcommand{\propositionname}{Proposition}}
\addto\captionsenglish{\renewcommand{\theoremname}{Theorem}}
\providecommand{\definitionname}{Definition}
\providecommand{\examplename}{Example}
\providecommand{\lemmaname}{Lemma}
\providecommand{\propositionname}{Proposition}
\providecommand{\theoremname}{Theorem}

\begin{document}

\title{Spooky effect in optimal OSPA estimation and how GOSPA solves it}

\author{Ángel F. García-Fernández\foreignlanguage{british}{$^{\star}$},
Lennart Svensson\foreignlanguage{british}{$^{\circ}$}\\
\foreignlanguage{british}{{\normalsize{}$^{\star}$}}{\normalsize{}Dept.
of Electrical Engineering and Electronics, University of Liverpool,
United Kingdom}\\
\foreignlanguage{british}{{\normalsize{}$^{\circ}$}}{\normalsize{}Dept.
of Electrical Engineering, Chalmers University of Technology, Sweden}\\
{\normalsize{}Emails: angel.garcia-fernandez@liverpool.ac.uk, lennart.svensson@chalmers.se}}

\maketitle
\thispagestyle{empty}
\begin{abstract}
In this paper, we show the spooky effect at a distance that arises
in optimal estimation of multiple targets with the optimal sub-pattern
assignment (OSPA) metric. This effect refers to the fact that if we
have several independent potential targets at distant locations, a
change in the probability of existence of one of them can completely
change the optimal estimation of the rest of the potential targets.
As opposed to OSPA, the generalised OSPA (GOSPA) metric ($\alpha=2$)
penalises localisation errors for properly detected targets, false
targets and missed targets. As a consequence, optimal GOSPA estimation
aims to lower the number of false and missed targets, as well as the
localisation error for properly detected targets, and avoids the spooky
effect. 
\end{abstract}


\section{Introduction}

Multiple target estimation is an inherent part of many applications
such as surveillance, self-driving vehicles, and air-traffic control
\cite{Blackman_book99,Petrovskaya09,Granstrom18b}. The special characteristic
of multiple target estimation is that it requires the estimation of
the number of targets, which is unknown, as well as their states. 

In a Bayesian paradigm, given some noisy observations of a random
variable of interest, all information about this variable is contained
in its posterior probability density function \cite{Sarkka_book13}.
Given the posterior and a cost function, optimal estimation is performed
by minimising the expected value of this cost function with respect
to the posterior \cite{Robert_book07,Kay_book93}. For example, for
random vectors of fixed dimensionality, if the cost function is the
square error, the optimal estimator, which is referred to as the minimum
mean square error estimator, is the posterior mean. 

In multi-target systems, the variable of interest can be represented
as a set of unknown cardinality and whose elements are the target
states \cite{Mahler_book07}. As in systems of fixed dimensionality,
developing optimal estimators for multi-target systems is important,
as they use all the information in the posterior density to provide
estimates with the smallest possible error. Before developing an optimal
multi-target estimator, a key aspect is the choice of a cost function
that measures errors in a suitable way and, therefore, yields desirable
properties for the optimal multi-target estimator. In this paper,
we analyse and discuss properties of optimal estimators based on multi-target
metrics, which we proceed to review. 

The Hausdorff metric is a general metric for sets and was proposed
to be used for sets of targets in \cite{Hoffman04}, but it is relatively
insensitive to differences in the number of targets \cite{Hoffman04}.
The Wasserstein metrics were originally used to measure the similarity
between probability distributions \cite{Barrio99,Villani_book09}
and were proposed for sets of targets in \cite{Hoffman04}. However,
they lack a physically consistent interpretation when the sets have
different cardinalities \cite{Schuhmacher08}. The optimal sub-pattern
assignment (OSPA) metric, which has parameters $p$ and $c$, was
firstly introduced to measure the similarity between distributions
of point processes \cite[page 669]{Schuhmacher08_b}. The OSPA metric
was proposed to be used for sets of targets in \cite{Schuhmacher08}
and has better properties for multi-target error evaluation than Hausdorff
and Wassertein metrics. The use of OSPA for optimal multiple target
estimation with known number of targets has been considered in \cite{Guerriero10,Baum15b,Baum12,Lipsa16,Crouse11d,Svensson11}.
With unknown number of targets, the use of unnormalised OSPA (UOSPA)
for multi-target estimation was proposed in \cite{Williams15}. The
cardinalized optimal linear assignment (COLA) metric was proposed
in the context of map estimation in robotics \cite{Barrios17}. COLA
corresponds to the UOSPA metric divided by $c$, which implies that
optimal estimators for COLA and UOSPA are the same.  

The generalised OSPA (GOSPA) metric \cite{Rahmathullah17} generalises
the UOSPA metric by adding a parameter $\alpha$ to adjust the cardinality
mismatch penalty. Importantly, if and only if $\alpha=2$, the GOSPA
metric can be written in terms of assignment sets, in which targets
can be left unassigned and only nearby targets are assigned to each
other \cite[Proposition 1]{Rahmathullah17}. In this case, GOSPA decomposes
into localisation errors for properly detected targets (assigned targets),
costs for missed targets and costs for false targets (unassigned targets).
Therefore, the GOSPA metric, contrary to OSPA and UOSPA, favours estimates
that locate detected targets well and keep the number of false and
missed targets to a minimum, as in traditional multiple target tracking
assessment methods \cite{Drummond92,Fridling91,Rothrock00,Mabbs93}.
For example, adding false targets to an estimate does not necessarily
increase the OSPA/UOSPA error, but it always increases GOSPA error
\cite[Example 2]{Rahmathullah17}. 

In this work, we show that optimal mean square OSPA and UOSPA estimators
produce an effect, which we refer to as spooky effect at a distance,
due to a similar effect in the cardinality probability hypothesis
density filter \cite{Franken09}. The spooky effect in optimal estimation
refers to the fact that a small change in the probability of existence
of one potential target can dramatically change the optimal estimation
of far-away independent potential targets. This is especially significant
with OSPA, as the appearance of a potential target with a small probability
of existence can trigger that all potential targets in the scene are
detected, even if their probabilities of existence are low. We also
show that the spooky effect is absent in optimal mean square GOSPA
($\alpha=2$) estimation. In this case, the optimization problem for
independent potential targets in distant regions can be separated
into local problems, such that the detection of a potential targets
depends only on its distribution, and not the distribution of the
rest of the targets. 

The rest of the paper is organised as follows. Section \ref{sec:Background}
reviews the considered metrics and the optimal estimation problem.
The spooky effect in optimal OSPA and UOSPA metric is explained in
Section \ref{sec:Spooky-effect}. Section \ref{sec:Spooky-effect}
also shows that optimal GOSPA estimation does not present spooky effect.
Finally, conclusions are drawn in Section \ref{sec:Conclusions}.

\section{Background\label{sec:Background}}

In this section, we review the OSPA, the UOSPA and the GOSPA metrics
and provide the conceptual solution to the optimal estimation problem. 

\subsection{Metrics}

We consider parameters $c>0$ and $1\leq p<\infty$. We also consider
$d\left(\cdot,\cdot\right)$ to be a metric in the single target space,
which is typically $\mathbb{R}^{n_{x}}$, and $d^{\left(c\right)}\left(\cdot,\cdot\right)=\min\left(d\left(\cdot,\cdot\right),c\right)$.
Let $\prod_{n}$ denote the set of all permutations of $\left\{ 1,...,n\right\} $
where $n\in\mathbb{N}$ and any element $\pi\in\prod_{n}$ can be
written as $\pi=\left(\pi\left(1\right),..,\pi\left(n\right)\right)$.
Also, let $X=\left\{ x_{1},...,x_{\left|X\right|}\right\} $ and $Y=\left\{ y_{1},...,y_{\left|Y\right|}\right\} $
denote two finite sets of single targets, with $\left|X\right|\leq\left|Y\right|$,
and $\left|X\right|$ being the cardinality of set $X$.
\begin{defn}
The OSPA metric between $X$ and $Y$ for $\left|Y\right|>0$ is \cite{Schuhmacher08_b,Schuhmacher08}
\begin{align*}
 & d_{p}^{\left(c\right)}\left(X,Y\right)\\
 & =\min_{\pi\in\prod_{\left|Y\right|}}\left(\frac{1}{\left|Y\right|}\sum_{i=1}^{\left|X\right|}d^{\left(c\right)}\left(x_{i},y_{\pi\left(i\right)}\right)^{p}+c^{p}\left(\left|Y\right|-\left|X\right|\right)\right)^{1/p}.
\end{align*}
For $\left|Y\right|=0$ and $\left|X\right|=0$, the OSPA metric is
$d_{p}^{\left(c\right)}\left(\emptyset,\emptyset\right)=0$.  $\boxempty$
\end{defn}
\smallskip{}
\begin{defn}
Given $0<\alpha\leq2$, the GOSPA metric between $X$ and $Y$  is
\cite{Rahmathullah17}
\begin{align*}
 & d_{p}^{\left(c,\alpha\right)}\left(X,Y\right)\\
 & =\min_{\pi\in\prod_{\left|Y\right|}}\left(\sum_{i=1}^{\left|X\right|}d^{\left(c\right)}\left(x_{i},y_{\pi\left(i\right)}\right)^{p}+\frac{c^{p}}{\alpha}\left(\left|Y\right|-\left|X\right|\right)\right)^{1/p}.\;\boxempty
\end{align*}
The differences with OSPA are the removal of the normalisation by
$\left|Y\right|$ and the additional parameter $\alpha$ to control
the cardinality mismatch penalty. The unnormalised OSPA (UOSPA) metric
corresponds to $d_{p}^{\left(c,1\right)}\left(X,Y\right)$. The key
property of the GOSPA metric is that, for $\alpha=2$, we can write
the metric in terms of assignment sets, which allows its decomposition
in terms of localisation error for properly detected targets, false
targets and missed targets. 
\end{defn}
\begin{prop}
\label{prop:GOSPA_alpha2}Let $\gamma$ be an assignment set between
$\left\{ 1,...,\left|X\right|\right\} $ and $\left\{ 1,...,\left|Y\right|\right\} $,
which meets $\gamma\subseteq\left\{ 1,...,\left|X\right|\right\} \times\left\{ 1,...,\left|Y\right|\right\} $,
$\left(i,j\right),\left(i,j'\right)\in\gamma\rightarrow j=j'$, and
$\left(i,j\right),\left(i',j\right)\in\gamma\rightarrow i=i'$. The
last two properties ensure that every $i$ and $j$ gets at most one
assignment. Then, the GOSPA metric ($\alpha=2$), can be written as
\cite[Prop. 1]{Rahmathullah17}
\begin{align}
 & d_{p}^{\left(c,2\right)}\left(X,Y\right)\nonumber \\
 & =\min_{\gamma\in\Gamma}\left(\sum_{\left(i,j\right)\in\gamma}d^{p}\left(x_{i},y_{j}\right)+\frac{c^{p}}{2}\left(\left|X\right|+\left|Y\right|-2\left|\gamma\right|\right)\right)^{1/p}\label{eq:GOSPA_alpha2}
\end{align}
where $\Gamma$ is the set of all possible $\gamma$.
\end{prop}
It should be noted that there is no cut-off parameter for $d\left(\cdot,\cdot\right)$
in (\ref{eq:GOSPA_alpha2}). The first term in (\ref{eq:GOSPA_alpha2})
represents the localisation errors for assigned targets (properly
detected ones) and the second term is the cost for the $\left|X\right|+\left|Y\right|-2\left|\gamma\right|$
unassigned targets, which includes missed and false targets. In the
rest of the paper, we refer to GOSPA with $\alpha=2$ simply as GOSPA.

\subsection{Optimal estimation}

In multiple target tracking, all information of interest about the
current set of targets is given by its multi-target density given
present and past measurements. This posterior density can be calculated
using the prediction and update steps of the Bayesian recursion \cite{Mahler_book07}.
In this work, we drop time indices and consider that the posterior
is  $f\left(\cdot\right)$.

In order to obtain optimal estimators, we consider minimum mean square
OSPA (MSOSPA), UOSPA (MSUOSPA) and GOSPA (MSGOSPA) errors with $p=2$.
It should be noted that minimising the MSOSPA, MSUOSPA and MSGOSPA
is equivalent to minimising the root MSOSPA, MSUOSPA and MSGOSPA,
which are themselves metrics for random finite sets \cite[Prop. 2]{Rahmathullah17}.
The optimal estimator in MSGOSPA sense (and analogously for OSPA and
UOSPA) is given by
\begin{align}
\hat{X}_{o} & =\underset{\hat{X}}{\arg\min}\,\mathrm{E}\left[\left(d_{2}^{\left(c,2\right)}\left(X,\hat{X}\right)\right)^{2}\right]\nonumber \\
 & =\underset{\hat{X}}{\arg\min}\int\left(d_{2}^{\left(c,2\right)}\left(X,\hat{X}\right)\right)^{2}f\left(X\right)\delta X\nonumber \\
 & =\underset{\hat{X}}{\arg\min}\sum_{n=0}^{\infty}\frac{1}{n!}\int\left(d_{2}^{\left(c,2\right)}\left(\left\{ x_{1},...,x_{n}\right\} ,\hat{X}\right)\right)^{2}\nonumber \\
 & \quad\quad\quad\quad\quad\times f\left(\left\{ x_{1},...,x_{n}\right\} \right)dx_{1:n}
\end{align}
where the integral corresponds to the set integral \cite{Mahler_book07}
and $x_{1:n}=\left(x_{1},...,x_{n}\right)$.

\section{Spooky effect in optimal multi-target estimators\label{sec:Spooky-effect}}

In this section, we analyse two simple scenarios in which we can calculate
the mean square errors for the metrics and the optimal multi-target
estimators analytically. This analysis provides important insights
into the behaviour of different optimal estimators. In particular,
we show that optimal estimators based on OSPA and UOSPA suffer from
the spooky effect. We also show that this effect appears in the marginal
multitarget estimator and joint multitarget estimators \cite{Mahler_book07}.
Importantly, optimal GOSPA estimation does not suffer from the spooky
effect.

In Section \ref{subsec:Set-up}, we explain the considered posterior
density. The resulting mean square errors, which are required to compute
the optimal estimators, for the different metrics are given in Section
\ref{subsec:Mean-square-errors}. The analysis when the posterior
has two Bernoulli components is given in Section \ref{subsec:Two-Bernoulli-components}.
The analysis when the posterior has an increasing number of Bernoulli
components is given in Section \ref{subsec:Increasing-Bernoullis}. 

\subsection{Posterior density\label{subsec:Set-up}}

Suppose the posterior $f\left(\cdot\right)$ is a multi-Bernoulli
density with $N$ Bernoulli components with known target locations
such that
\begin{align}
f\left(X\right) & =\sum_{X_{1}\uplus...\uplus X_{N}=X}\prod_{i=1}^{N}f_{i}\left(X_{i}\right),\label{eq:MB_density}
\end{align}
where $\uplus$ denotes the disjoint union and the density of the
$i$-th Bernoulli component is
\begin{align}
f_{i}\left(X\right) & =\begin{cases}
1-r_{i} & X=\emptyset\\
r_{i}\delta\left(x-\overline{x}_{i}\right) & X=\left\{ x\right\} \\
0 & \mathrm{otherwise}
\end{cases}\label{eq:Bernoulli_density}
\end{align}
where $r_{i}$ is the probability of existence of the $i$-th Bernoulli,
$\delta\left(\cdot\right)$ is a Dirac delta and $\overline{x}_{i}$
is the location of the $i$-th Bernoulli component. In addition, we
consider that all the Bernoulli components are sufficiently far from
each other $d\left(\overline{x}_{i},\overline{x}_{j}\right)>c$ for
$i\neq j$. Note that the summation in (\ref{eq:MB_density}) is
taken over all mutually disjoint, and possibly empty, sets $X_{1},...,X_{N}$
whose union is $X$. 

While the results in this section hold for $d\left(\overline{x}_{i},\overline{x}_{j}\right)>c$,
the spooky effect becomes clearer if potential targets (Bernoulli
components) are quite far from each other $d\left(\overline{x}_{i},\overline{x}_{j}\right)\gg c$.
We would like to remark that we consider Bernoulli densities with
known locations, as in this case, the mean square errors admit closed-form
formulas.  

\subsection{Mean square errors and optimal estimators\label{subsec:Mean-square-errors}}

We proceed to obtain the mean square errors for the OSPA/UOSPA/GOSPA
metrics, as a preliminary step to obtain the optimal estimators. Due
to the fact that the single target densities of the Bernoulli components
are Dirac deltas, the optimal estimate for all metrics must be a subset
of $\left\{ \overline{x}_{1},...,\overline{x}_{N}\right\} $. Any
other choice increases the error and is therefore non-optimal, see
proof in Appendix \ref{sec:AppendixA}. We parameterise the possible
estimates within this set by a vector $\hat{e}_{1:N}=\left[\hat{e}_{1},...,\hat{e}_{N}\right]^{T}$
where $\hat{e}_{i}=1$ if the $i$-th Bernoulli component is detected
and zero otherwise. That is, given this parameterisation, the estimated
set is given by
\begin{align*}
\hat{X} & =\left\{ \overline{x}_{i}:\hat{e}_{i}=1,\,i\in\left\{ 1,...,N\right\} \right\} .
\end{align*}
\begin{lem}
The mean square errors for the different metrics, the estimate parameterised
by $\hat{e}_{1:N}$ and the posterior density (\ref{eq:MB_density})
are
\begin{align}
\mathrm{MSGOSPA} & =\frac{c^{2}}{2}\sum_{i=1}^{N}\left[r_{i}\left(1-\hat{e}_{i}\right)+\left(1-r_{i}\right)\hat{e}_{i}\right]\label{eq:MSGOSPA}\\
\mathrm{MSUOSPA} & =c^{2}\left[\sum_{n=0}^{N}\rho\left(n\right)\max\left(n,\hat{n}\right)-\sum_{i=1}^{N}\hat{e}_{i}r_{i}\right]\label{eq:MSUOSPA}
\end{align}
and
\begin{align}
 & \mathrm{MSOSPA}\nonumber \\
 & \quad=\begin{cases}
c^{2}\left(1-\sum_{i=1}^{N}\left[\hat{e}_{i}r_{i}\sum_{n=0}^{N-1}\frac{\rho_{-i}\left(n\right)}{\max\left(n+1,\hat{n}\right)}\right]\right) & \hat{n}>0\\
c^{2}\left(1-\rho\left(0\right)\right) & \hat{n}=0
\end{cases}\label{eq:MSOSPA}
\end{align}
where
\begin{align}
\hat{n} & =\sum_{j=1}^{N}\hat{e}_{j}\label{eq:n_hat}
\end{align}
is the number of detected targets, $\rho\left(\cdot\right)$ is the
cardinality distribution of the multi-Bernoulli density \cite[page 102]{Mahler_book14},
and $\rho_{-i}\left(\cdot\right)$ is the cardinality distribution
of the multi-Bernoulli density without the $i$-th Bernoulli component. 
\end{lem}
This lemma is proved in Appendix \ref{sec:AppendixB}.  The optimal
estimates can be obtained by minimising (\ref{eq:MSGOSPA})-(\ref{eq:MSOSPA})
with respect to $\hat{e}_{1:N}$. In the MSGOSPA error (\ref{eq:MSGOSPA}),
there is a sum over all Bernoulli components and each term of the
sum is the MSGOSPA error for the corresponding Bernoulli component.
It then follows that the optimal MSGOSPA estimator admits a closed-form
solution
\begin{align}
\hat{e}_{i} & =\begin{cases}
1 & r_{i}>0.5\\
0 & \mathrm{otherwise}.
\end{cases}\label{eq:GOSPA_optimal_estimator}
\end{align}
It is relevant to highlight that each potential target is detected
based only on its probability of existence. The probabilities of existence
of other targets do not affect the estimate of a target. 

On the contrary, the errors for UOSPA and OSPA cannot be written as
the sum of the errors for each Bernoulli component. This implies that
the estimation problem is not disentangled. Instead, the estimate
of a Bernoulli component depends on what happens in distant parts
of the state space, which creates the spooky effect at a distance
\cite{Franken09}.  The optimal estimator for UOSPA and OSPA does
not have a simple expression as in GOSPA, which is given by (\ref{eq:GOSPA_optimal_estimator}),
but it can be obtained by evaluating the errors for all possible values
of $\hat{e}_{1:N}$.

\subsection{Two Bernoulli components\label{subsec:Two-Bernoulli-components}}

\begin{figure}
\begin{centering}
\includegraphics[scale=0.5]{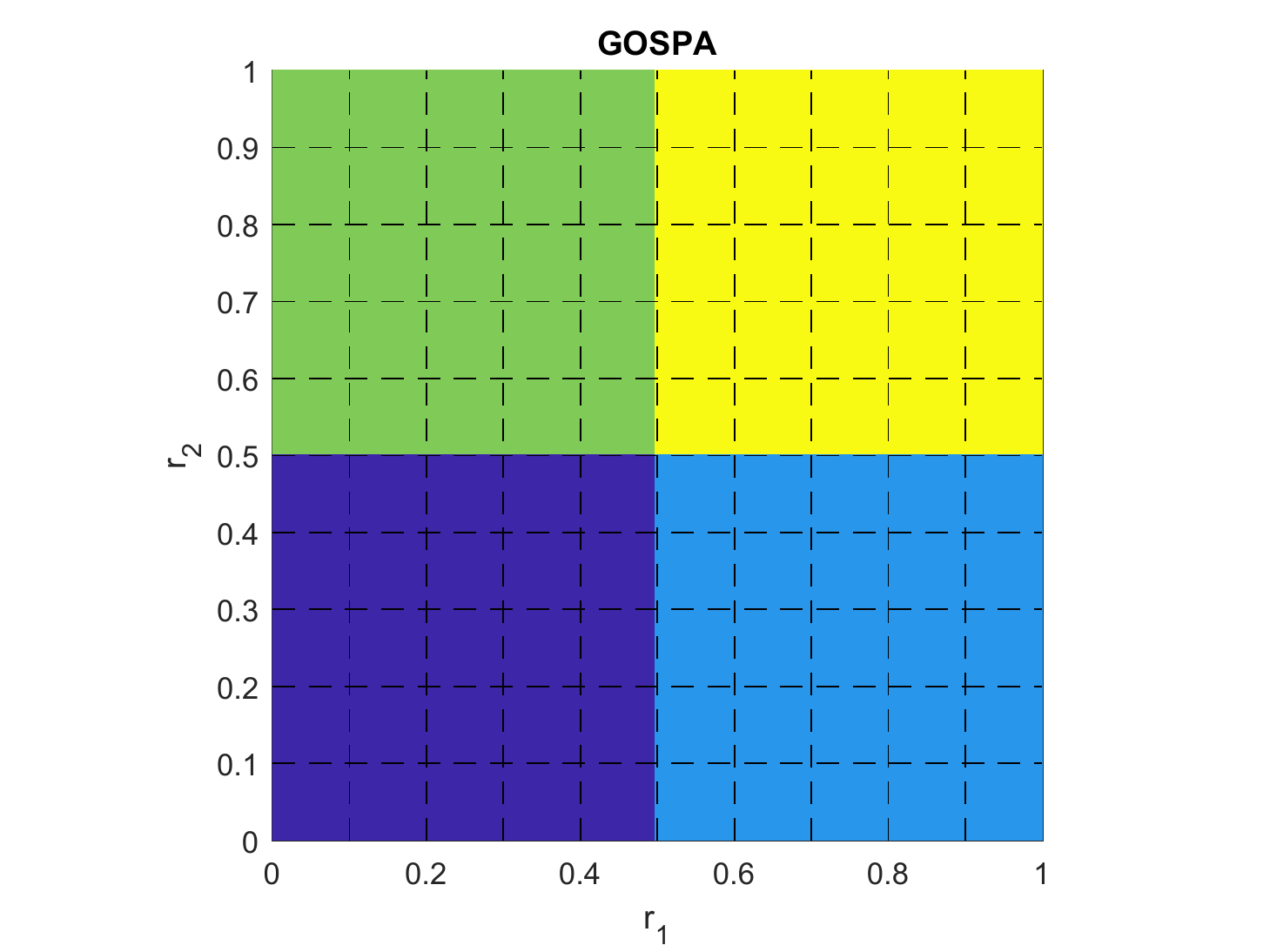}
\par\end{centering}
\begin{centering}
\includegraphics[scale=0.5]{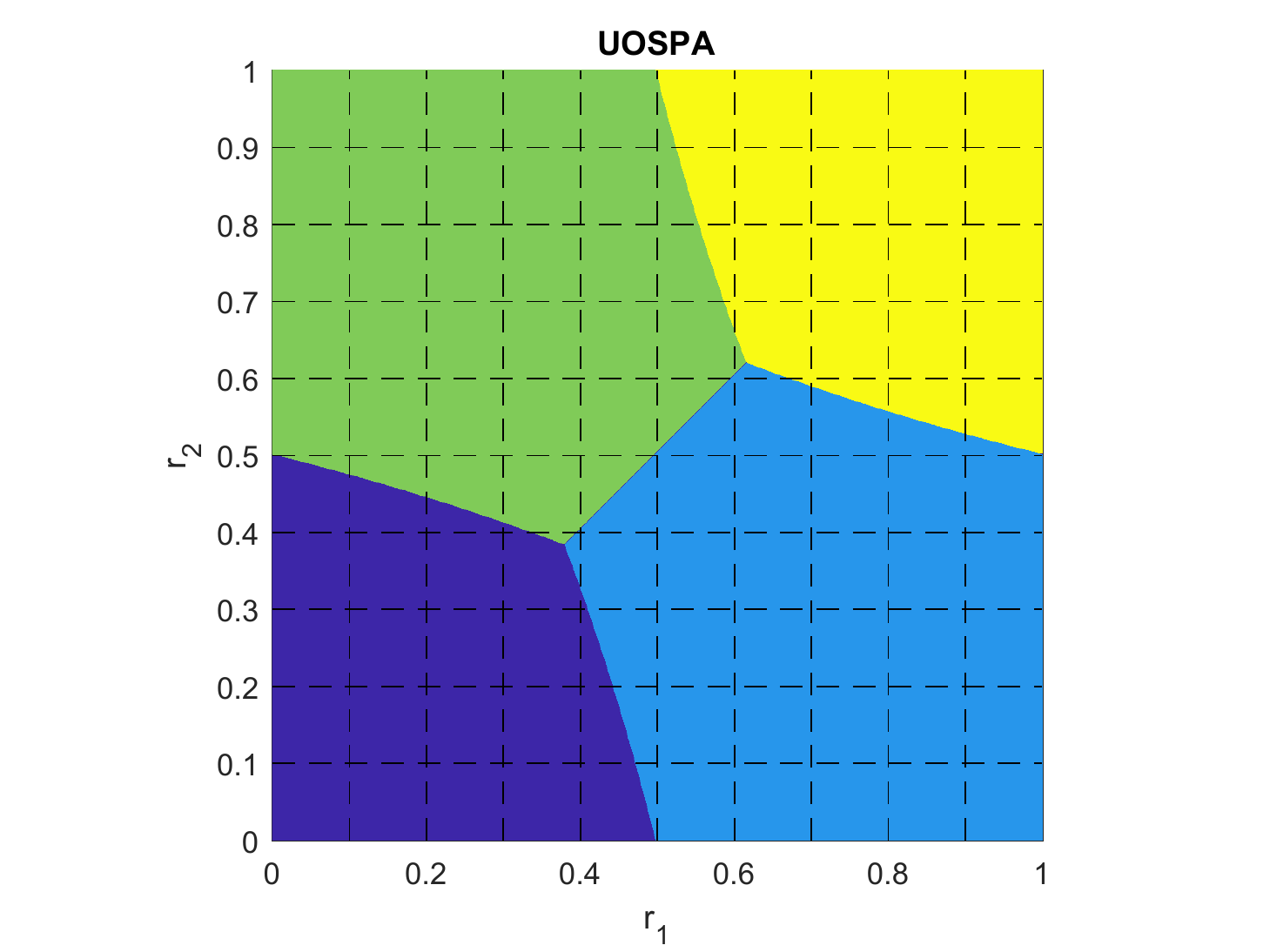}
\par\end{centering}
\begin{centering}
\includegraphics[scale=0.5]{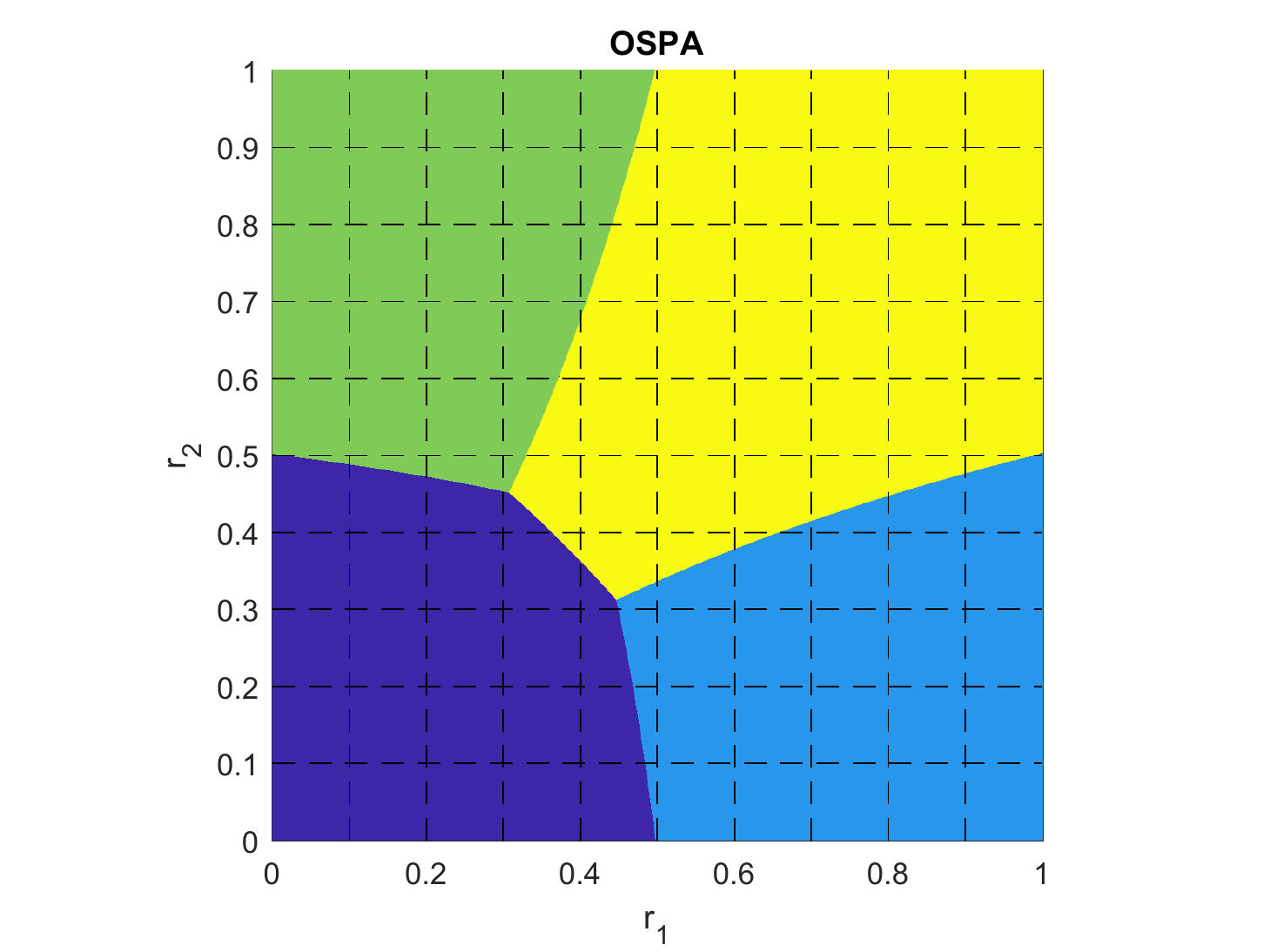}
\par\end{centering}
\caption{\label{fig:Decision-regions_metrics}Decision regions for the optimal
estimators based on GOSPA, UOSPA and OSPA against the existence probabilities
of two Bernoulli components. Dark blue: no target is estimated. Light
blue: only target 1 is estimated. Green: only target 2 is estimated.
Yellow: both targets are estimated. The only metric whose optimal
estimator is free from spooky effect is GOSPA ($\alpha=2$). }
\end{figure}

In this section, we analyse the optimal estimates for GOSPA, OSPA
and UOSPA in a case where there are two Bernoulli components, $N=2$.
The corresponding optimal estimators, obtained as indicated in Section
\ref{subsec:Mean-square-errors}, are shown in Figure \ref{fig:Decision-regions_metrics}.
According to this figure, for all optimal estimators, if the probability
of existence of one Bernoulli component is zero, a target estimate
is optimally reported if its probability of existence is higher than
0.5. However, when we add an independent, far away potential target
($r_{2}>0$), only GOSPA is able to preserve this property. The rest
of the optimal estimators show spooky effect at a distance, as the
optimal estimator in one area is influenced by independent events
in far-away regions. This can lead to counter intuitive results, as
illustrated in the following example. 

\begin{figure}
\begin{centering}
\includegraphics[scale=0.5]{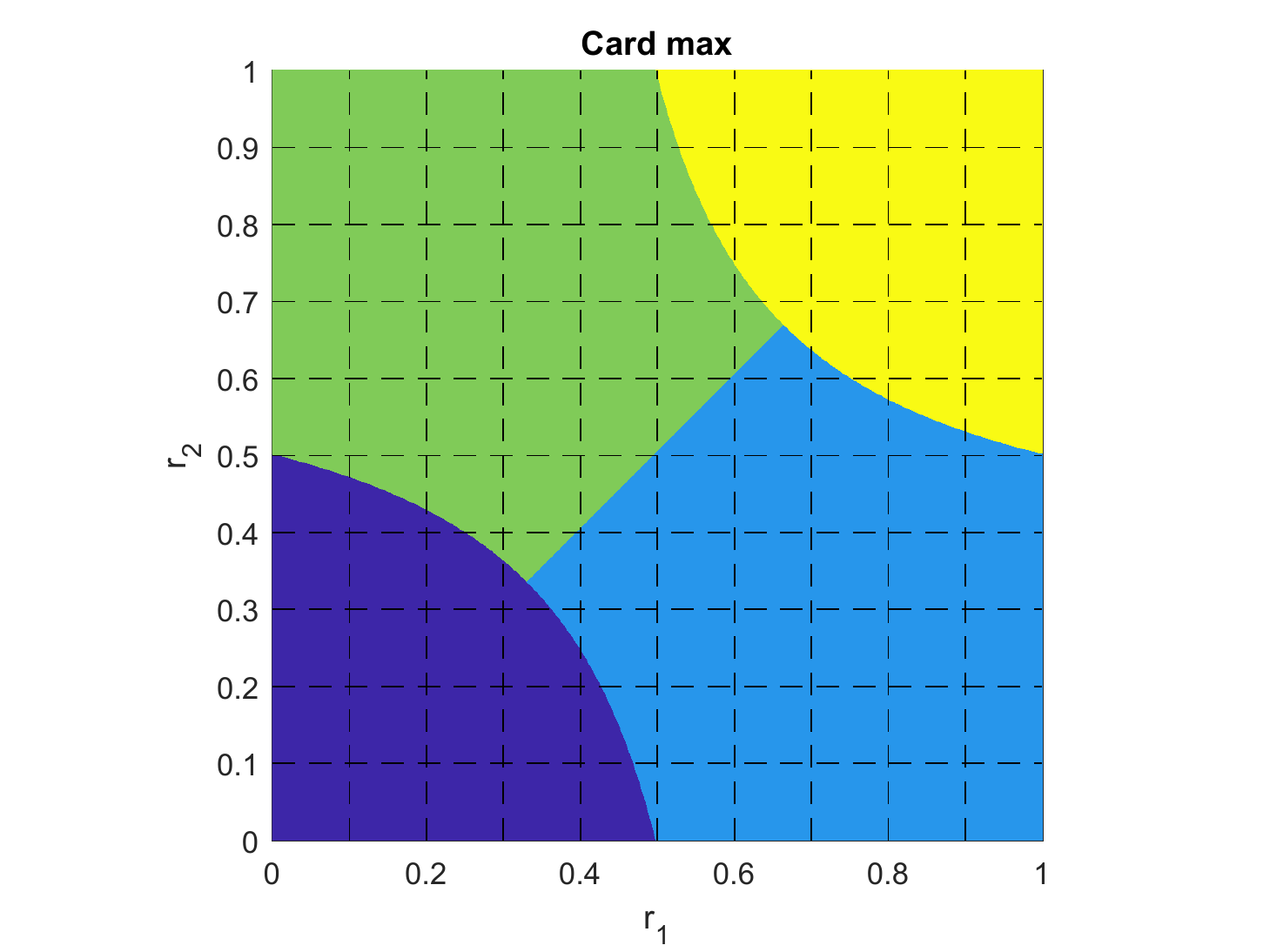}
\par\end{centering}
\begin{centering}
\includegraphics[scale=0.5]{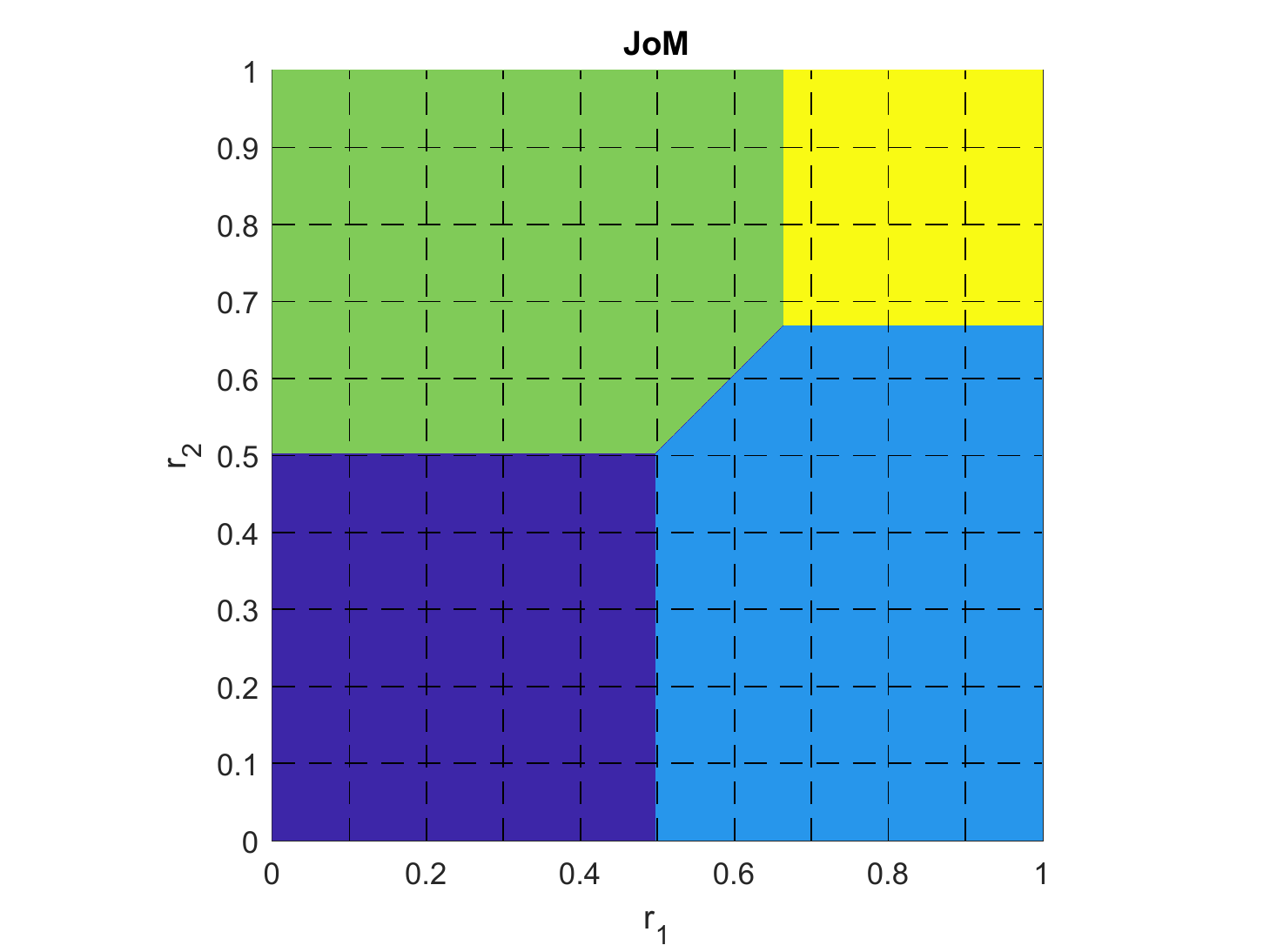}
\par\end{centering}
\caption{\label{fig:Decision-regions-other_estimators}Decision regions for
the marginal multitarget estimator, which coincide with the regions
for the estimator that maximises the cardinality distribution first,
(top) and the joint multitarget estimator (bottom) against the existence
probabilities of two Bernoulli components. These estimators show spooky
effect.}
\end{figure}

\begin{example}
Let us consider that there are two potential targets: Bernoulli component
1 in Madrid and Bernoulli component 2 in Liverpool. These potential
targets are independent of each other, but are being tracked by the
same system. The probability of existences are $r_{1}=0.4$ and $r_{2}=0.4$.
Therefore, the optimal OSPA estimator reports two targets, see Figure
\ref{fig:Decision-regions_metrics}. We analyse the following cases
\end{example}
\begin{itemize}
\item Case 1: We receive a measurement from the potential target in Liverpool,
such that $r_{2}$ increases to $r_{2}=0.9$. The measurement from
the potential target in Liverpool conveys no information whatsoever
on the potential target in Madrid, and $r_{1}$ is not modified. However,
now, the optimal OSPA estimator only reports target 2. In other words,
an increase in the probability of existence of one potential target
can actually make that the other potential target is no longer being
reported, even if they are independent events in far-away regions.
Apart from the spooky effect, it is also interesting to observe that
optimal OSPA estimation has a counter intuitive behaviour with respect
to the estimation of the total number of targets in the scene. That
is, before taking the measurement, the optimal OSPA estimator reports
the two targets. Once we receive the measurement, the probability
that there are two targets actually increases, but the optimal OSPA
estimator chooses to drop one of the previously reported targets. 
\item Case 2: We receive a measurement from the potential target in Liverpool,
such that $r_{2}$ decreases to $r_{2}=0.3$, which does not affect
$r_{1}$. In this case, using an optimal OSPA estimator, both potential
targets are no longer detected. As in the previous case, the optimal
estimation of a potential target changes depending on a independent
event in a distant region. $\boxempty$
\end{itemize}
Situations with spooky effect also arise in optimal UOSPA estimation.
It is also relevant to analyse if the spooky effect at a distance
is observed in other types of multi-target estimators, not based on
metrics, commonly used in the literature. We consider the marginal
multitarget estimator, the joint multitarget estimators (JoM) \cite{Mahler_book07,Baser15,Baser16}
and the estimator that first maximises the cardinality distribution
and then reports the targets with largest existence probability, as
in \cite{Vo13}. The marginal multitarget estimator and JoM are not
defined when the Bernoulli densities include Dirac deltas. Nevertheless,
we apply them by considering Gaussian distributions with a covariance
matrix $\sigma^{2}I$, with small $\sigma^{2}$, instead of Dirac
deltas in (\ref{eq:Bernoulli_density}). We set the JoM parameter
$c$ ($c$ as defined in \cite[Sec. 14.5]{Mahler_book07}) to $1/\mathcal{N}\left(0;0,\sigma^{2}I\right)$,
which removes the dependency of the JoM on $\sigma^{2}$. The resulting
decision regions of the optimal estimators are shown in Figure \ref{fig:Decision-regions-other_estimators}.
All of them show spooky effect, as the optimal estimator for two Bernoulli
components does not make decisions independently for each Bernoulli
component. 

\subsection{Increasing number of Bernoulli components\label{subsec:Increasing-Bernoullis}}

We proceed to analyse the effect of increasing the number $N$ of
Bernoulli components on the optimal estimators based on UOSPA, OSPA
and GOSPA. We consider that the existence probabilities of all Bernoulli
components are the same, $r_{i}=r$ for $i=1,...,N$ and that all
Bernoulli components are far from each other.

In this case, the mean square UOSPA and OSPA errors can be simplified
as follows
\begin{align}
\mathrm{MSUOSPA} & =c^{2}\left[\sum_{n=0}^{N}\rho\left(n\right)\max\left(n,\hat{n}\right)-\hat{n}r\right]\nonumber \\
\mathrm{MSOSPA} & =\begin{cases}
c^{2}\left(1-\hat{n}r\sum_{n=0}^{N-1}\frac{\rho_{-1}\left(n\right)}{\max\left(n+1,\hat{n}\right)}\right) & \hat{n}>0\\
c^{2}\left(1-\rho\left(0\right)\right) & \hat{n}=0
\end{cases}\label{eq:MSOSPA_same_r}
\end{align}
where we have applied that $\rho_{-1}\left(\cdot\right)=\rho_{-i}\left(\cdot\right)$
for all $i$ as all the probabilities of existence are alike. Note
that the above mean square errors depend on the estimated number of
targets, not the individual target estimates $\hat{e}_{i}$ $i=1,...,N$,
due to the fact that existence probabilities are alike.

As shown in Appendix \ref{sec:AppendixC}, the optimal estimator for
MSOSPA detects $\hat{n}_{0}$ targets, with
\begin{align}
\hat{n}_{0} & =\begin{cases}
N & \left(1-r\right)^{N}<r\\
0 & \mathrm{otherwise}.
\end{cases}\label{eq:optimal_estimator_MSOSPA}
\end{align}
That is, the optimal OSPA estimator either detects 0 or all the targets
depending on the inequality in the previous equation, which depends
on the number of Bernoulli components and the probability of existence.

\begin{figure}
\begin{centering}
\subfloat[]{\begin{centering}
\includegraphics[scale=0.6]{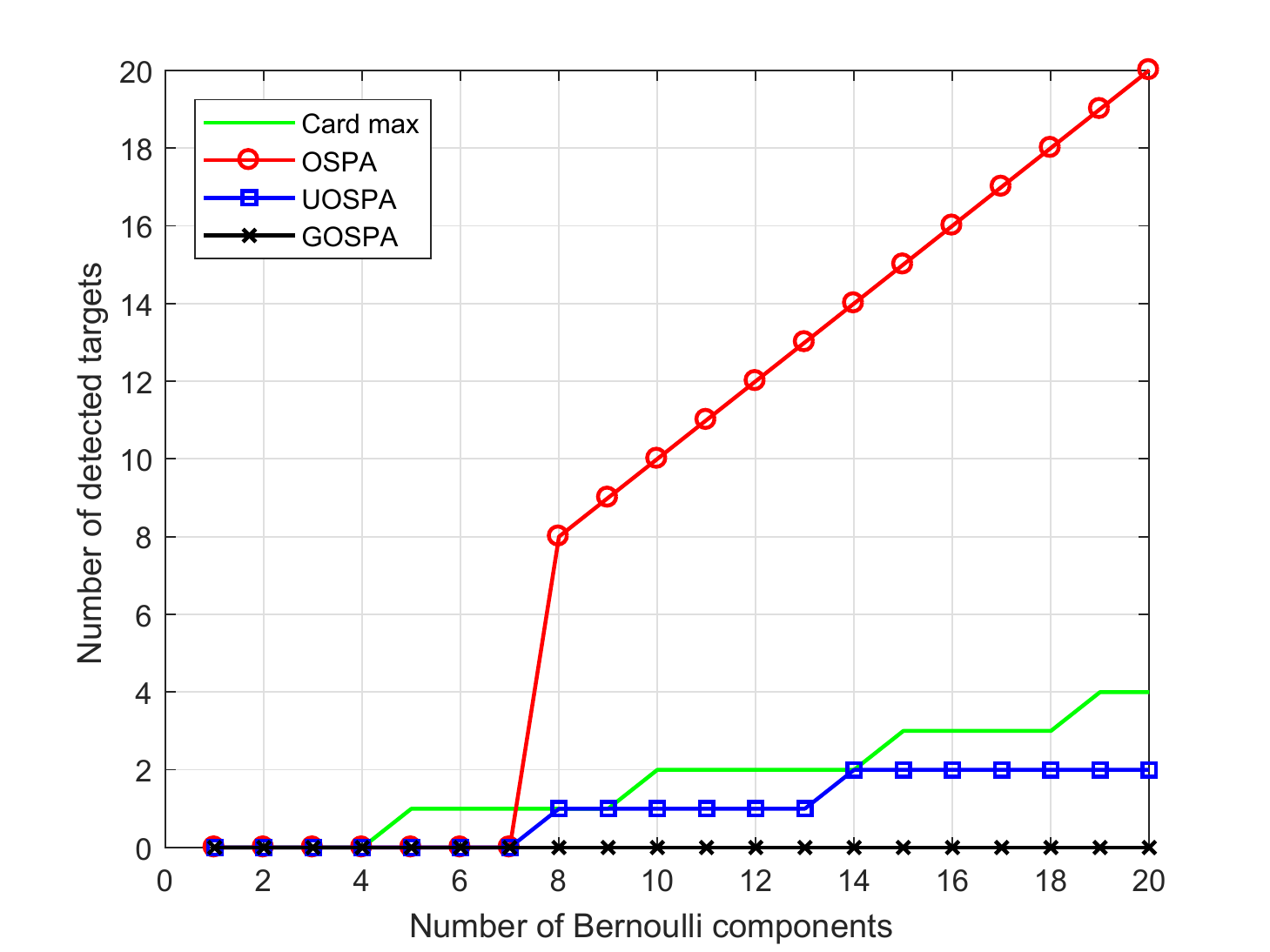}
\par\end{centering}
}
\par\end{centering}
\begin{centering}
\subfloat[]{\begin{centering}
\includegraphics[scale=0.6]{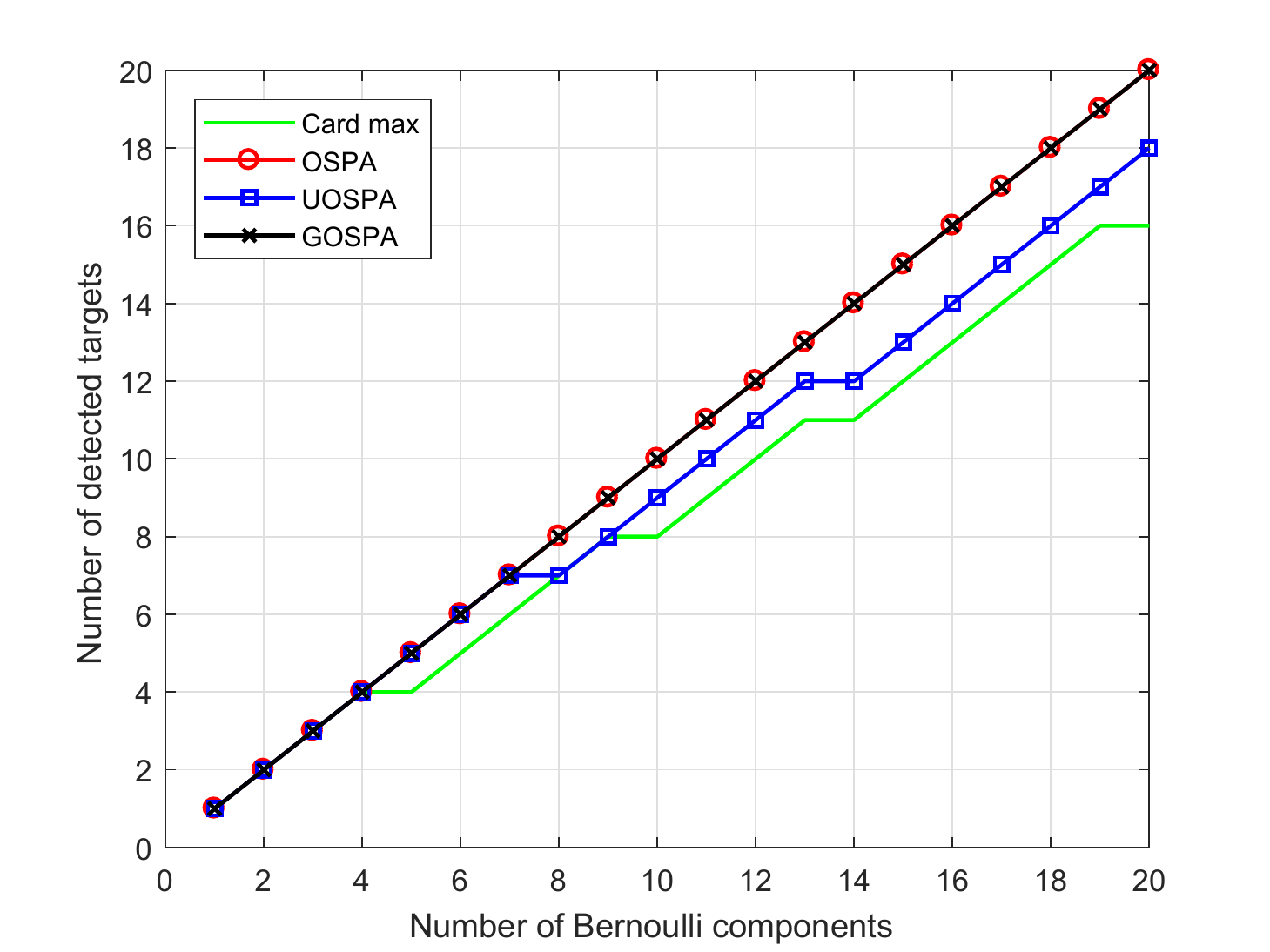}
\par\end{centering}
}
\par\end{centering}
\caption{\label{fig:Optimal-number_detections_far_away}Optimal number of detected
targets against the number of Bernoulli components at far distance
for the different metrics for (a) $r=0.2$ (b) $r=0.8$. For (a) OSPA
optimally detects all targets if there are more than 7 targets, otherwise
it does not detect any. UOSPA optimally detects 0 targets for less
than 8 Bernoulli components and then adds target detections as there
are more targets. GOSPA does not detect any targets as all of them
are far away with a low probability of existence. In (b), OSPA optimally
detects all targets. UOSPA detects all targets if there are less than
8 Bernoulli components but then misses some of the targets for a higher
number of Bernoulli components. GOSPA detects all the targets as all
of them are far away with a high probability of existence. OSPA and
UOSPA show spooky effect in optimal estimation as if we add independent
Bernoulli components in far away regions, it affects the target reports
in other regions. }
\end{figure}

We plot the number of optimally detected targets against the number
of Bernoulli components for $r=0.2$ and $r=0.8$ in Figure \ref{fig:Optimal-number_detections_far_away}.
GOSPA detects a target if its probability of existence is higher than
0.5, see (\ref{eq:GOSPA_optimal_estimator}), independently of the
rest of the Bernoulli components. For OSPA, we can see from (\ref{eq:optimal_estimator_MSOSPA}),
that, for increasing $N$, there is a point at which the optimal number
of detected targets jumps directly from 0 to $N$. This means that
adding an independent Bernoulli component in a far away region, even
with a very small probability, can make the estimator go from no detections
at all to detect all possible targets. This is a counterintuitive
behaviour of a multi-target estimator in standard applications. For
$r=0.8$, OSPA detects all targets for $N\geq1$, as GOSPA. 

In the case of UOSPA, for $r=0.2$, the optimal number of detected
targets increases with $N$. However, UOSPA adds them one by one,
which is more reasonable than the optimal OSPA estimator. For $r=0.8$,
the optimal UOSPA estimator does not detect all targets as $N$ increases
and it drops targets one by one. For example, for $N=8$, it does
not detect one target and, for $N=14$, it does not detect two targets.
Optimal UOSPA estimation has the advantage over optimal OSPA estimation
that it avoids the all-or-nothing estimation effect, but it still
shows spooky effect. Adding a far away Bernoulli component can either
create the detection of a previously non-detected target or can remove
a detection of a previously detected targets. It is also interesting
to note that, as all targets have the same existence probability,
an optimal UOSPA estimator can select any combination of them, with
the right number of targets, to be reported. In addition, we can see
that the estimator that reports the number of targets that maximises
the cardinality distribution is more similar to UOSPA than to OSPA
or GOSPA.

\section{Conclusions\label{sec:Conclusions}}

This paper has presented the spooky effect at a distance in optimal
OSPA and UOSPA estimation. This effect is non desired in standard
multi-target estimation, as the spooky effect in the CPHD filter \cite{Franken09}.
The spooky effect can be avoided intrinsically by a selection of a
suitable multi-target metric, the GOSPA metric, and its corresponding
optimal estimator. 

We think that the spooky effect in OSPA and UOSPA  is a reason to
support the use of GOSPA in standard multitarget tracking problems,
where one aims to localise targets well and lower the number of false
and missed targets. An additional benefit of GOSPA is that one can
report the error decomposition\footnote{Matlab code of the GOSPA metric and its decomposition is available
at https://github.com/abusajana/GOSPA

A short video that explains GOSPA is available at https://www.youtube.com/watch?v=M79GTTytvCM } along with the overall metric value to show greater insights into
the performances of different algorithms, as done in \cite{Xia17}.

We would also like to remark that there can be applications where
the spooky effect, either of optimal OSPA/UOSPA estimation or the
CPHD filter, can actually be beneficial. If in a certain application,
it is more important to determine the overall number of objects well,
compared to localising objects with sufficient accuracy and lowering
missed and false objects, OSPA, UOSPA or GOSPA with general $\alpha$
may be the best choice of metric for this problem. In this case, the
spooky effect should be promoted by optimal estimators. Nevertheless,
in conventional multiple target tracking, the spooky effect is arguably
both undesirable and counterintuitive. 

\appendices{}

\section{\label{sec:AppendixA}}

In this appendix, we show that the optimal estimate must be a subset
of $\left\{ \overline{x}_{1},...,\overline{x}_{N}\right\} $ for the
problem formulation in Section \ref{sec:Spooky-effect}. We first
prove it for mean square GOSPA, with general $\alpha$. We have
\begin{align*}
\mathrm{MSGOSPA_{\alpha}} & =\sum_{n=0}^{\infty}\frac{1}{n!}\int\left(d_{2}^{\left(c,\alpha\right)}\left(\left\{ x_{1},...,x_{n}\right\} ,\hat{X}\right)\right)^{2}\\
 & \quad\times f\left(\left\{ x_{1},...,x_{n}\right\} \right)dx_{1:n}\\
 & =\left[\prod_{i=1}^{N}\left(1-r_{i}\right)\right]\sum_{n=0}^{N}\frac{1}{n!}\sum_{1\leq i_{1}\neq...\neq i_{n}\leq N}\\
 & \,\left(d_{2}^{\left(c,\alpha\right)}\left(\hat{X},\left\{ \overline{x}_{i_{1}},...,\overline{x}_{i_{n}}\right\} \right)\right)^{2}\prod_{j=1}^{n}\frac{r_{i_{j}}}{1-r_{i_{j}}}\\
 & =\left[\prod_{i=1}^{N}\left(1-r_{i}\right)\right]\sum_{n=0}^{N}\sum_{\left\{ i_{1},...,i_{n}\right\} \subseteq\left\{ 1,...,N\right\} }\\
 & \,\left(d_{2}^{\left(c,\alpha\right)}\left(\hat{X},\left\{ \overline{x}_{i_{1}},...,\overline{x}_{i_{n}}\right\} \right)\right)^{2}\prod_{j=1}^{n}\frac{r_{i_{j}}}{1-r_{i_{j}}}\\
 & =\sum_{I\subseteq\left\{ 1,...,N\right\} }p\left(I\right)\left(d_{2}^{\left(c,\alpha\right)}\left(\hat{X},\overline{X}\left(I\right)\right)\right)^{2}
\end{align*}
where we have used the expression of the multi-Bernoulli density in
\cite[Eq. (4.127)]{Mahler_book14} and
\begin{align*}
p\left(\left\{ i_{1},...,i_{n}\right\} \right) & =\left[\prod_{i=1}^{N}\left(1-r_{i}\right)\right]\prod_{j=1}^{n}\frac{r_{i_{j}}}{1-r_{i_{j}}},\\
\overline{X}\left(\left\{ i_{1},...,i_{n}\right\} \right) & =\left\{ \overline{x}_{i_{1}},...,\overline{x}_{i_{n}}\right\} .
\end{align*}

If $\hat{X}$ is a subset of $\overline{X}$, then
\begin{align*}
d_{2}^{\left(c,\alpha\right)}\left(\hat{X},\overline{X}\left(I\right)\right) & =\frac{c^{2}}{\alpha}\left[\max\left(\left|\hat{X}\right|,\left|\overline{X}\left(I\right)\right|\right)\right.\\
 & \quad\left.-\min\left(\left|\hat{X}\right|,\left|\overline{X}\left(I\right)\right|\right)\right]
\end{align*}
for all $I$. That is, for the optimal permutation, the terms that
include $d^{\left(c\right)}\left(\cdot,\cdot\right)$ in the metric
are always zero. 

On the contrary, if $\hat{X}$ is not a subset of $\overline{X}$,
then
\begin{align*}
d_{2}^{\left(c,\alpha\right)}\left(\hat{X},\overline{X}\left(I\right)\right) & >\frac{c^{2}}{\alpha}\left[\max\left(\left|\hat{X}\right|,\left|\overline{X}\left(I\right)\right|\right)\right.\\
 & \quad\left.-\min\left(\left|\hat{X}\right|,\left|\overline{X}\left(I\right)\right|\right)\right]
\end{align*}
for at least one $I\subseteq\left\{ 1,...,N\right\} $ because one
of the terms $\sum_{i=1}^{\left|X\right|}d^{\left(c\right)}\left(x_{i},y_{\pi\left(i\right)}\right)^{2}$
in the optimal permutation must be higher than zero. Therefore, the
optimal MSGOSPA estimate for any $\alpha$ must be a subset of $\overline{X}$.
Similar arguments apply to OSPA.

\section{\label{sec:AppendixB}}

In this appendix, we prove the expressions of the mean square errors
for GOSPA, UOSPA and OSPA, which are given by (\ref{eq:MSGOSPA}),
(\ref{eq:MSUOSPA}) and (\ref{eq:MSOSPA}).

\subsection{GOSPA}

We use variable $e_{i}=1$ to denote the case that the $i$-th Bernoulli
component has an existing target and $e_{i}=0$, otherwise. Therefore,
the probability of the event $e_{1:N}=\left[e_{1},...,e_{N}\right]^{T}$
is
\begin{align*}
p\left(e_{1:N}\right) & =\prod_{i=1}^{N}\left[\left(1-r_{i}\right)\left(1-e_{i}\right)+r_{i}e_{i}\right].
\end{align*}

For each pair $e_{1:N}$ and $\hat{e}_{1:N}$, it is straightforward
to obtain that the square GOSPA error is
\begin{align*}
\mathrm{SGOSPA} & =\frac{c^{2}}{2}\sum_{i=1}^{N}\left[e_{i}\left(1-\hat{e}_{i}\right)+\left(1-e_{i}\right)\hat{e}_{i}\right].
\end{align*}
That is, each Bernoulli component contributes with an error $\frac{c^{2}}{2}$
if the target is not detected and exists or if the target is detected
and it does not exist. Otherwise, the error for each Bernoulli component
is zero.

Considering that the probability of existence of the $i$-th Bernoulli
component is $r_{i}$, see (\ref{eq:Bernoulli_density}) , the resulting
$\mathrm{MSGOSPA}$ error is given in (\ref{eq:MSGOSPA}). 

\subsection{UOSPA}

The number of targets in the ground truth is
\begin{align*}
n & =\sum_{i=1}^{N}e_{i}.
\end{align*}
The number of targets in the estimate is $\hat{n}$, see (\ref{eq:n_hat}).
For each pair $e_{1:N}$ and $\hat{e}_{1:N}$, it is direct to obtain
that the square UOSPA error is
\begin{align}
\mathrm{SUOSPA} & =c^{2}\left(\max\left(n,\hat{n}\right)-\sum_{i=1}^{N}\hat{e}_{i}e_{i}\right)\label{eq:square_UOSPA_appendix}
\end{align}
where the term $\sum_{i=1}^{N}\hat{e}_{i}e_{i}$ represents the number
of properly detected targets. In this case, a properly detected target
implies that the target and the estimate are in the same location.
It should be noted that, each properly detected target is penalised
with a zero error, and each of the rest of the targets in the largest
set is penalised with an error $c^{2}$. 

The mean square UOSPA error is then
\begin{align*}
\mathrm{MSUOSPA} & =c^{2}\sum_{e_{1:N}}p\left(e_{1:n}\right)\left(\max\left(n,\hat{n}\right)-\sum_{i=1}^{N}\hat{e}_{i}e_{i}\right)\\
 & =c^{2}\left[\sum_{e_{1:N}}p\left(e_{1:n}\right)\max\left(n,\hat{n}\right)-\sum_{i=1}^{N}\hat{e}_{i}r_{i}\right]\\
 & =c^{2}\left[\sum_{n=0}^{N}\rho\left(n\right)\max\left(n,\hat{n}\right)-\sum_{i=1}^{N}\hat{e}_{i}r_{i}\right]
\end{align*}
where $\rho\left(n\right)$ represents the cardinality distribution
of a multi-Bernoulli density \cite[page 102]{Mahler_book14}. 

\subsection{OSPA}

For each pair $e_{1},...,e_{N}$ and $\hat{e}_{1},...,\hat{e}_{N}$,
the square OSPA error is
\begin{align*}
\mathrm{SOSPA} & =\begin{cases}
0 & n=0,\,\hat{n}=0\\
c^{2}\left(1-\frac{\sum_{i=1}^{N}\hat{e}_{i}e_{i}}{\max\left(n,\hat{n}\right)}\right) & \mathrm{otherwise}
\end{cases}
\end{align*}
which corresponds to the square UOSPA error, see (\ref{eq:square_UOSPA_appendix}),
normalised by $\max\left(n,\hat{n}\right)$, for $n\neq0,\,\hat{n}\neq0$. 

The mean square OSPA error is then
\begin{align}
\mathrm{MSOSPA} & =\begin{cases}
c^{2}\left(1-\sum_{e_{1:n}}p\left(e_{1:n}\right)\frac{\sum_{i=1}^{N}\hat{e}_{i}e_{i}}{\max\left(n,\hat{n}\right)}\right) & \hat{n}>0\\
c^{2}\left(1-\rho\left(0\right)\right) & \hat{n}=0
\end{cases}\label{eq:MSOSPA_append1}
\end{align}

We proceed to simplify the term
\begin{align}
\sum_{e_{1:n}}p\left(e_{1:n}\right)\frac{\sum_{i=1}^{N}\hat{e}_{i}e_{i}}{\max\left(n,\hat{n}\right)} & =\sum_{i=1}^{N}\hat{e}_{i}\mathrm{E}\left[\frac{e_{i}}{\max\left(n,\hat{n}\right)}\right]\label{eq:MSOSPA_append2}
\end{align}
We first note that the total number of targets in the ground truth
can be written as $n=e_{i}+n_{-i}$ where $n_{-i}$ represents the
number of elements targets apart from the Bernoulli component $i$.
Due to the fact that existences are independent in the Bernoulli components,
we can write
\begin{align*}
p\left(e_{i},n_{-i}\right) & =p\left(e_{i}\right)\rho_{-i}\left(n_{-i}\right)
\end{align*}
where $\rho_{-i}\left(\cdot\right)$ represents the cardinality distribution
of all Bernoulli components except the $i$-th one. This cardinality
is known as it is the cardinality of a multi-Bernoulli distribution
\cite[page 102]{Mahler_book14}. 

Then, we have
\begin{align}
 & \mathrm{E}_{e_{i},n}\left[\frac{e_{i}}{\max\left(n,\hat{n}\right)}\right]\nonumber \\
 & \quad=\mathrm{E}_{e_{i},n_{-i}}\left[\frac{e_{i}}{\max\left(n_{-i}+e_{i},\hat{n}\right)}\right]\nonumber \\
 & \quad=\sum_{e_{i}=0}^{1}\sum_{n_{-i}=0}^{N-1}p\left(e_{i}\right)\rho_{-i}\left(n_{-i}\right)\frac{e_{i}}{\max\left(n_{-i}+e_{i},\hat{n}\right)}\nonumber \\
 & \quad=r_{i}\sum_{n=0}^{N-1}\frac{\rho_{-i}\left(n\right)}{\max\left(n+1,\hat{n}\right)}.\label{eq:MSOSPA_append3}
\end{align}
Finally, substituting (\ref{eq:MSOSPA_append2}) and (\ref{eq:MSOSPA_append3})
into (\ref{eq:MSOSPA_append1}), we obtain (\ref{eq:MSOSPA}).

\subsection{GOSPA with general $\alpha$}

For completeness, in this appendix, we write the expression for mean
square GOSPA error for general $\alpha$. 

For each pair $e_{1:N}$ and $\hat{e}_{1:N}$, the square GOSPA error,
general $\alpha$, is

\begin{align*}
\mathrm{SGOSPA}_{\alpha} & =\frac{c^{2}}{\alpha}\left|n-\hat{n}\right|+c^{2}\left(\min\left(n,\hat{n}\right)-\sum_{i=1}^{N}\hat{e}_{i}e_{i}\right).
\end{align*}
As before, the term $\sum_{i=1}^{N}\hat{e}_{i}e_{i}$ represents the
number of properly detected targets. These targets are penalised with
an error 0. Each of the rest of the targets in the smaller set is
penalised with an error $c^{2}$. Finally, the difference in cardinality
is penalised multiplied by a factor $\frac{c^{2}}{\alpha}$. Note
that $\left|n-\hat{n}\right|=\max\left(n,\hat{n}\right)-\min\left(n,\hat{n}\right)$,
so for $\alpha=1$, we recover the UOSPA error, as required.

The mean square GOSPA error is then
\begin{align*}
 & \mathrm{MSGOSPA}_{\alpha}\\
 & \,=\sum_{n=0}^{N}\left[\frac{c^{2}}{\alpha}\left|n-\hat{n}\right|+c^{2}\min\left(n,\hat{n}\right)\right]\rho\left(n\right)-c^{2}\sum_{i=1}^{N}\hat{e}_{i}r_{i}
\end{align*}

\section{\label{sec:AppendixC}}

In this appendix, we show that the optimal number of detected targets
using OSPA in the scenario in Section \ref{subsec:Increasing-Bernoullis}
is given by (\ref{eq:optimal_estimator_MSOSPA}). 

First, we calculate the MSOSPA error for $\hat{n}=N$ using Eq. (\ref{eq:MSOSPA_same_r}).
We have
\begin{align*}
\mathrm{MSOSPA}\left(N\right) & =c^{2}\left(1-Nr\sum_{n=0}^{N-1}\frac{\rho_{-1}\left(n\right)}{\max\left(n+1,N\right)}\right)\\
 & =c^{2}\left(1-Nr\sum_{n=0}^{N-1}\frac{\rho_{-1}\left(n\right)}{N}\right)\\
 & =c^{2}\left(1-r\sum_{n=0}^{N-1}\rho_{-1}\left(n\right)\right)\\
 & =c^{2}\left(1-r\right)
\end{align*}
where in the last step we have used that the cardinality distribution
of the multi-Bernoulli with $N-1$ Bernoulli components sums to one
over $n=0,...,N-1$. 

In addition, for $\hat{n}=0$, we can write the MSOSPA in terms of
the probability of existence to yield
\begin{align*}
\mathrm{MSOSPA}\left(0\right) & =c^{2}\left[1-\left(1-r\right)^{N}\right].
\end{align*}
 Therefore, it is better to detect $N$ targets instead of 0 targets
if
\begin{align*}
\left(1-r\right)^{N} & <r.
\end{align*}

The next part of the proof consists of showing that
\begin{align*}
\mathrm{MSOSPA}\left(\hat{n}\right) & >\mathrm{MSOSPA}\left(N\right)
\end{align*}
for $0<\hat{n}<N$, which means that the optimal estimate has either
zero or $N$ targets. 

First, for $0\leq n<N$ and $0<\hat{n}<N$, it holds that
\begin{align}
\hat{n} & \leq\max\left(n+1,\hat{n}\right)\nonumber \\
\frac{1}{\max\left(n+1,\hat{n}\right)} & \leq\frac{1}{\hat{n}}\nonumber \\
\frac{\rho_{-1}\left(n\right)}{\max\left(n+1,\hat{n}\right)} & \leq\frac{\rho_{-1}\left(n\right)}{\hat{n}}.\label{eq:Append_inequality_OSPA}
\end{align}

For $n=N-1$, we have a strict inequality in (\ref{eq:Append_inequality_OSPA})
since $N>\hat{n}$ when $0<\hat{n}<N$. We proceed to sum over $n=0$
to $N-1$ in both sides of (\ref{eq:Append_inequality_OSPA}). Since
we have a strict inequality in at least one term, the summation also
yields a strict inequality
\begin{align*}
\sum_{n=0}^{N-1}\frac{\rho_{-1}\left(n\right)}{\max\left(n+1,\hat{n}\right)} & <\sum_{n=0}^{N-1}\frac{\rho_{-1}\left(n\right)}{\hat{n}}\\
1-\hat{n}r\sum_{n=0}^{N-1}\frac{\rho_{-1}\left(n\right)}{\max\left(n+1,\hat{n}\right)} & >1-\hat{n}r\sum_{n=0}^{N-1}\frac{\rho_{-1}\left(n\right)}{\hat{n}}\\
\mathrm{MSOSPA}\left(\hat{n}\right) & >\mathrm{MSOSPA}\left(N\right)\quad0<\hat{n}<N
\end{align*}
which finishes the proof of (\ref{eq:optimal_estimator_MSOSPA}). 

\bibliographystyle{IEEEtran}
\bibliography{7E__Trabajo_Angel_Referencias_Referencias}

\end{document}